# Rare-earth doped yttrium silicate (Y$_2$SiO$_5$) thin films grown by chemical vapour deposition for quantum technologies

Suma Al-Hunaishi[1], Anna Blin[1], Nao Harada[1], Pauline Perrin[1], Philippe Goldner[1], Diana Serrano[1*] and Alexandre Tallaire[1#]

[1] IRCP, CNRS, PSL Research University, 11 rue Pierre et Marie Curie, 75005, Paris, France

* Corresponding author: diana.serrano@chimieparistech.psl.eu

# Corresponding author: alexandre.tallaire@chimieparistech.psl.eu

**Abstract**

Yttrium orthosilicate (Y$_2$SiO$_5$ - YSO) is one of the most promising crystals to host rare-earth (RE) ions for quantum technologies applications. In this matrix, they indeed exhibit narrow optical and spin linewidths that can be exploited to develop quantum memories or quantum information processing capabilities. In this paper, we propose a new method to grow RE doped silicate thin films on silicon wafers based on direct liquid injection chemical vapour deposition (DLI-CVD). We optimize the deposition and annealing conditions to achieve formation of the high temperature X2-YSO phase. The phase purity and crystalline quality of the films are assessed by evaluating the optical properties of Eu$^{3+}$ ions embedded in this oxide matrix. In view of the results, we discuss the possible phase formation mechanisms, and the potential of this new wafer-compatible form of YSO for quantum technologies applications.

1. Introduction

Quantum technologies aim to exploit the peculiarities of quantum physics to generate real-world tangible applications. They are bound to have a significant societal impact with possible breakthrough achievements in information processing, communications, or sensing. To be successful though, many technological building blocks encompassing a wide range of disciplines will have to be established. Those include material sciences which play a significant role by providing "quantum-grade" materials through advanced synthesis and processing methods. Among the candidate material systems currently under development, rare-earth (RE) ions possess tremendous potential [1]. When doped in a crystalline matrix, they behave as optically active centres with transitions that span the entire visible range, as well as the IR one (in the telecom wavelength for Er$^{3+}$ in particular). These transitions occur within 4$f$ electrons that are shielded from their environment by closed shells. As a result, they are exceptionally narrow for a solid-state system, reaching sub-kilohertz optical linewidths [2,3]. At the same time, RE ions present optically addressable long-lived electron and/or nuclear spin states [4,5]. This unique combination of narrow optical and spin linewidths is exploited for the implementation of quantum memory protocols [6] and quantum processing schemes [7].

Bulk RE crystals are the workhorse in this field and in particular RE-doped yttrium orthosilicate ($RE^{3+}:Y_2SiO_5$ hereinafter referred to as $RE^{3+}$:YSO) has so far provided unmatched performance in terms of optical and spin coherence times [2,4,5,8]. High-quality YSO crystals are typically grown from the melt using crystal pulling techniques at temperatures as high as 2000 °C [9,10] and then oriented and sliced. Their practical integration into useful and versatile quantum devices would however be facilitated if nanoscale materials such as nanoparticles and thin films were available. For instance, efficient coupling to tuneable fibre cavities requires active materials fitting small mode volumes, typically subwavelength [11]. In addition, nanoscale materials could be assembled forming hybrid structures providing advanced functionalities and scaling-up opportunities. Finally, they offer great prospects due to their lower thermal budget and precursor consumption during synthesis which constitutes a cost-effective alternative when isotopically purified material is needed [3].

The bottom-up synthesis of "quantum-grade" RE nanomaterials has been mostly focused on binary oxides. For example, $Eu^{3+}$-doped $Y_2O_3$ nanoparticles have been produced and reached record optical and spin coherence lifetimes at the nanoscale [12,13]. In addition, $Y_2O_3$ thin films deposited on silicon have been obtained by different deposition techniques, including molecular beam epitaxy (MBE) [14,15], atomic layer deposition (ALD) [16] or chemical vapour deposition (CVD) [17,18]. Recently, spin coherence times ($T_2$) in the millisecond range combined with a 3 kHz optical dephasing rate have been reported with an MBE-grown $Y_2O_3$ thin film doped with 25 ppm $Er^{3+}$ [19]. Some examples of micro and nanoscale YSO prepared by bottom-up synthesis can be found in the literature [20,21]. Among them, YSO films have been considered for environmental barrier coatings in aeronautics due to their advantageous thermal and mechanical properties [22], or as phosphors [23] and scintillators [24] when doped with RE ions. However, they have not been extensively studied in view of assessing their utility for quantum applications. This is most likely due to the difficulty in finding appropriate methods to synthesize this ternary compound with sufficient crystalline quality and phase purity.

In this paper, we propose a new approach to grow thin films of RE-doped YSO on silicon wafers, a versatile and scalable platform, based on direct liquid injection chemical vapour deposition (DLI-CVD) [25]. By selecting adequate precursors and feeding them into our liquid injection system we successfully obtained coatings with a controlled thickness, variable composition and phase depending on the deposition and subsequent annealing conditions. In particular, we identified a set of parameters allowing for the obtention of the high-temperature phase (X2) of YSO. We then evaluated the optical properties of Eu$^{3+}$ ions embedded in this matrix and assessed their performance by benchmarking them with high-quality bulk X2-YSO crystals obtained by conventional crystal growth methods. The relevance of this thin-film fabrication approach and its potential in the context of quantum technologies is discussed.

This paper is dedicated to the memory of François Auzel (1938-2023), a widely recognized pioneer in the field of modern luminescence. One of us (PG) had the honour to work in his group for a few years at the CNET laboratories at the end of the 90's. He is mostly known for his discovery in 1966, of the conversion of infrared to visible light through energy transfer[26], which is a major topic of current research. However, as an extremely creative scientist, François Auzel explored many other ideas in a broad range of materials, including rare earth doped thin films[27], which is the focus of this work. Interested in materials, physics, theory, and experimental work, he had an impressive mastery of the field that, combined with ever-new ideas, shaped his scientific journey. He is deeply missed as a scientist and mentor.

## 2. Experimental details

The yttria silicate films were deposited by DLI-CVD in a home-made reactor described in previous papers [17,18]. It consisted of a cold wall chamber in which 2-inch (111)-oriented silicon wafers, used as substrates, were placed onto a resistive heater. The yttrium and europium precursors were Y(tmhd)$_3$ and Eu(tmhd)$_3$ (where "tmhd" accounts for tetra-methyl-heptane-dionate) that are solid at room temperature, with 4N and 3N purity respectively. The precursors were dissolved in mesitylene to a concentration of 10 mmol/L for Y and 0.18 mmol/L for Eu, corresponding to a Y/Eu atomic ratio of 1.8 % and stored in the same pressurized canister. The selected silicon precursor, liquid DADBS (di-t-butoxydiacetoxysilane), was reported to offer good thermal stability and easy handling [28]. This precursor was also diluted in mesitylene to a concentration of 23 mmol/L. The liquids were flown into a 2-injection head vaporization box heated to 170 °C (*Vapbox 1500* from *Kemstream*). They were pulse-sprayed, and flash-evaporated and the resulting vapours were brought to the reacting chamber using N$_2$ as a carrier gas. Molecular oxygen was also added during growth to a flow ratio of about 1/3 with respect to nitrogen. By varying the opening time and frequency of the injectors it was possible to precisely control the actual incoming molar flow into the chamber and thus adjust the Y to Si ratio. The yttrium/europium mixture molar flow was fixed to 8 µmol/min while different silica molar flows of 4, 8 and 12 µmol/min were chosen. The deposition was performed at a temperature of about 750 °C as measured by a thermocouple placed underneath the heater and a pressure of 13 mbar. The films were later annealed at atmospheric pressure and at 1200 °C in air for 2 hours. The growth conditions are summarized in Table 1.

| Sample | Y(tmhd)$_3$ molar flow (µmol/min) | DADBS molar flow (µmol/min) | Eu/Y ratio (%at.) | Y/Si ratio (%at.) |
|---|---|---|---|---|
| A | 8 | 4 | 1.8 | 2 |
| B | 8 | 8 | 1.8 | 1 |
| C | 8 | 12 | 1.8 | 0.66 |

Table 1: Deposition conditions. The Eu/Y and Y/Si ratios are derived from the molar flows used during CVD growth.

The film's thicknesses after growth were estimated to be in the range 200-250 nm using spectroscopic ellipsometry with a *Woolam iSE* set-up. The fit was performed with the *Complete Ease* software using a thin interfacial SiO$_2$ layer and a Cauchy equation well adapted for transparent oxides of unknown composition. Differential interference contrast microscopy images were recorded with an *Olympus* microscope equipped with a Nomarski prism allowing enhancing the surface topography of the samples. The photoluminescence (PL) spectra were measured at room temperature using a *Renishaw InVia Raman* system operating with a 532 nm laser for the excitation. The system allows performing hyperspectral PL maps across large areas with a given step size. Lifetime measurements of Eu$^{3+}$ ions in the YSO thin film (sample A) were carried out under green excitation at 532 nm using a home-built confocal microscope equipped with a high numerical aperture objective (NA = 0.95). The sample was mounted on an XYZ positioning system (P-*611.3 NanoCube*). PL decays were recorded by modulating the excitation using a mechanical chopper. The setup is equipped with a photon-counting module (*COUNT-10C*), a spectrometer and a CCD camera (*DU401A-BVF*), providing spectral information about the collected fluorescence signal.

The optical properties of a bulk Eu$^{3+}$:YSO reference sample (0.1 %at., cut and polished into 500 µm slabs) were measured using a tuneable optical parametric oscillator (OPO) pumped by a Nd$^{3+}$ YAG Q-switched laser (*Ekspla NT342B-SH* with 6 ns pulse length and 10 Hz repetition rate). Spectra were recorded by a spectrometer (*Acton SP2300*) equipped with 300 grooves/mm holographic grating and an ICCD camera (*Princeton Instruments*), while a photomultiplier tube (*Hamamatsu EMI 9658B*) was used as detector for lifetime measurements.

3. Results

PL emission spectra were collected at room temperature for all samples in Table 1 and analysed, providing direct information about their crystalline quality and phase. Fig. 1 (left) shows the PL spectra for thin-film samples A (Fig. 1(b)), B (Fig. 1(c)) and C (Fig. 1(d)) immediately after DLI-CVD growth compared to the PL spectrum of $Eu^{3+}$ ions in a reference $Y_2O_3$ thin film (Fig. 1(a)) [17]. It can be observed that the sample grown with the lowest Si precursor (sample A, Fig. 1(b)) is comparable to the reference spectrum. The dominant emission peak at 612 nm, attributed to the $^5D_0 \rightarrow {}^7F_2$ $Eu^{3+}$ transition in $Y_2O_3$, can be also vaguely recognized in sample B (Y/Si=1, Fig. 1(c)). In contrast, no clear $Eu^{3+}$ emission peaks are observed in sample C, with the highest Si content (Y/Si = 0.66, Fig. 1(d)), suggesting the absence of extended crystalline phases in this film. Instead, a broad luminescence centred at around 585 nm is observed, which origin is unknown. These results indicate that the Y/Si ratio determines the dominant phase in the as-grown films, varying from amorphous for higher Si contents to cubic $Y_2O_3$ when increasing the Y/Si ratio. We also note that no crystalline silicate phases were directly obtained after deposition.

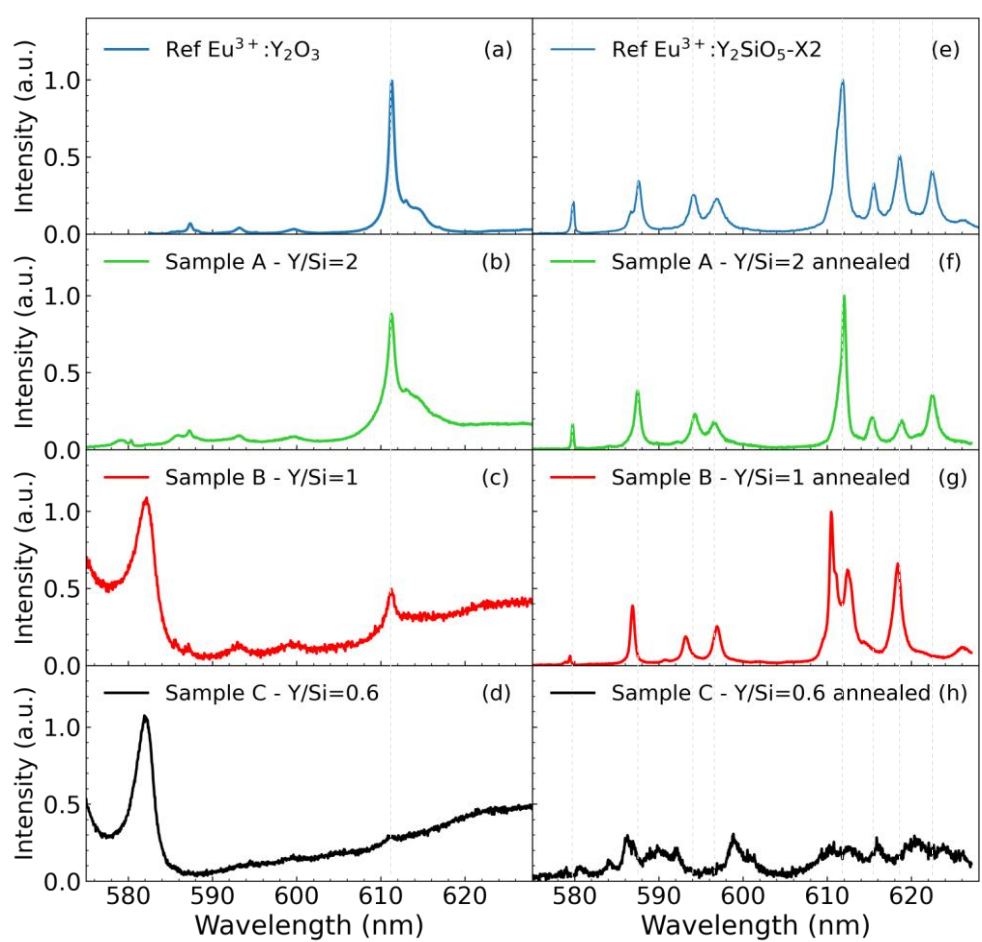

Fig. 1. (a)-(d). Room temperature PL spectra obtained under 532 nm excitation of as-grown films with different Y/Si ratio. The emission spectrum of a $Y_2O_3$:$Eu^{3+}$ CVD film (2 % at.) [17] recorded in the same conditions is given for comparison. (e)-(g) Room temperature PL spectra obtained under 532 nm excitation of the same films after annealing them in air at 1200 °C for 2 h. The emission spectrum of a bulk X2-YSO: $Eu^{3+}$ crystal (0.1 % at.) recorded in the same conditions is given for comparison.

A significant change in the emission spectra is observed after annealing the films for 2 h in air at 1200°C (Fig. 1, right column). Interestingly, for the film grown with a Y/Si ratio of 2 (sample A, Fig. 1(f)) the annealing treatment leads to the formation of the high- temperature YSO phase, referred to as X2 [29]. This is confirmed by the position of the $Eu^{3+}$ emission lines, similar to those observed in the bulk $Eu^{3+}$: X2-YSO (0.1 % at.) single crystal (Fig. 1(e)). It is however worth noting that there are some small differences in the spectral line's relative intensities between sample A (Fig. 1(f)) and the bulk reference (Fig. 1(e)). In X2-YSO, $Eu^{3+}$ ions replace $Y^{3+}$ ions in two different point-symmetry sites, yielding two different emission spectra. Fig. 2 displays these emission spectra recorded from the bulk reference crystal under selective laser excitation at 10 K (spectra i and ii). The room temperature spectra of the bulk (iii) and thin film (iv) samples are also given, showing contributions from both sites. The difference between film and bulk crystal spectra could be due to variations in the relative contribution from the two sites resulting from a different crystal orientation with respect to the excitation field.

After annealing, sample B (Y/Si = 1) yields a different emission spectrum from sample A, shown in Fig. 1(g), tentatively attributed to the disilicate phase: $\beta$-$Y_2Si_2O_7$. This is suggested by the appearance of emission lines identical to those reported for this phase in nanocrystals [30]. Appearance of orthosilicate and disilicate phases for samples A and B respectively is consistent with the Y/Si ratio set during synthesis. Finally, for the film with the highest Si amount (Fig. 1(h)), we could not detect any PL emissions indicating that the film remained mostly amorphous even after annealing.

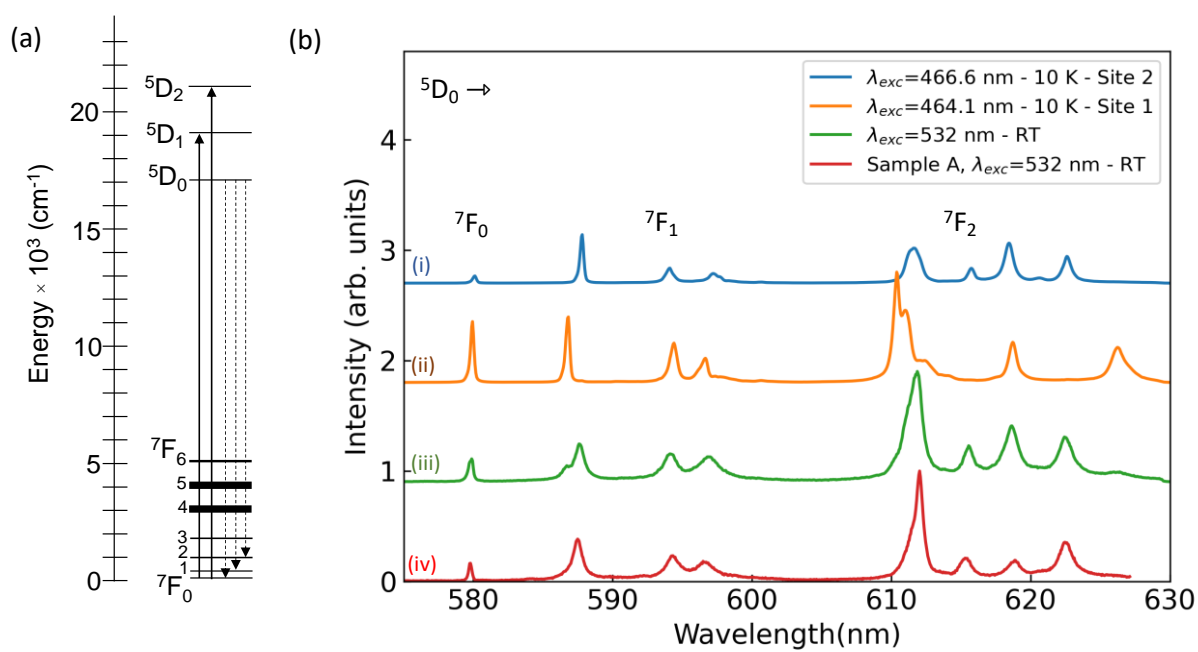

Fig. 2. (a) $Eu^{3+}$ energy level scheme. Straight lines indicate the excited transitions at 532 nm ($^5D_1$) and at 464.1 and 466.6 nm ($^5D_2$). Dashed lines indicate the observed emission lines. (b) Emission spectra from the reference bulk sample measured at 10 K and showing characteristic emission peaks from $Eu^{3+}$ ions in site 1 (ii) and site 2

(i) in X2-YSO. The lines are attributed as follows: $^5D_0 \rightarrow {}^7F_0$ (580 nm), $^5D_0 \rightarrow {}^7F_1$ (587, 594 and 596 nm), and $^5D_0 \rightarrow {}^7F_2$ (611, 615, 618, 622 and 626 nm). (iii) Room temperature emission spectrum of the reference bulk sample. (iv) Room temperature emission spectrum of sample A after annealing at 1200 °C.

Despite having successfully formed the desired X2-YSO phase after annealing for sample A, a more detailed investigation indicates that crystallisation was limited to specific areas at the surface of the film. Indeed, the microscope optical images show a smooth and uniform surface immediately after growth (Fig. 3a) while after annealing at 1200°C, flower-like round-shape patterns were formed (Fig. 3b).

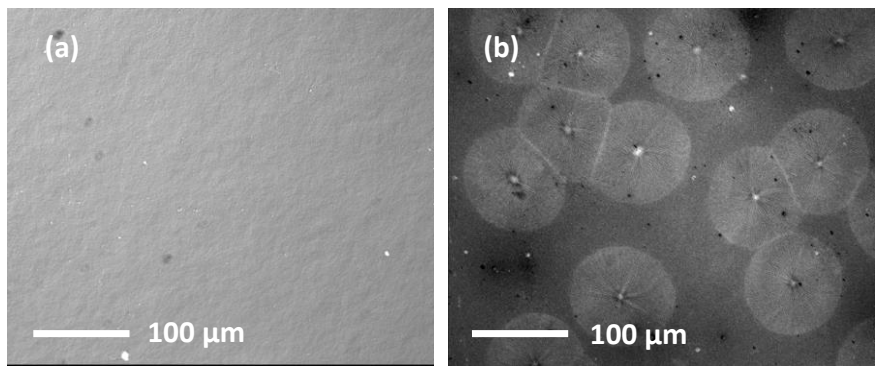

Fig. 3. Differential interference contrast microscopy images of the surface of sample A (Y/Si=2) acquired (a) before and (b) after annealing.

To further assess the spatial localisation and total amount of the crystallized X2-YSO phase in sample A, we recorded PL maps over several mm² areas. An example is shown in Fig. 4(b), representing the PL intensity at 623 nm, that was chosen because it corresponds to an $Eu^{3+}$ emission line only attributed to the X2-YSO phase. This emission line is highlighted in red in the spectrum of Fig. 4(a) that is similar to that of Fig. 1f. Many spots can be observed in the map, indicating formation of the X2-YSO phase at multiple places. In addition, the comparison with PL maps plotted using wavelengths common to other phases such as 612 nm, indicate the absence of parasitic phases in this sample. The total amount of X2-YSO is estimated to about 15 % while dark areas in the PL map, accounting for 85 % of the total surface, remained poorly crystallized or amorphous. To further refine this analysis, a zoom into one of the spots is shown in Fig. 4(c). One can recognize flower-shape patterns with a diameter of about 100 µm like those observed in Fig. 3(b), confirming a good match between those patterns and the formation of the X2-YSO phase after annealing.

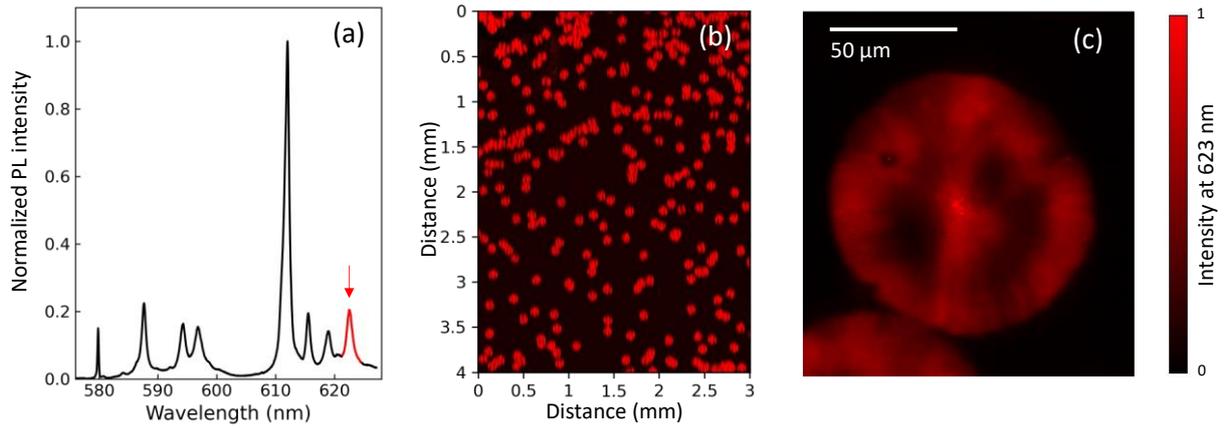

Fig. 4. (a) Emission spectrum from $Eu^{3+}$ in X2-YSO obtained from one of the crystallized spots in sample A after annealing. The emission line at 623 nm (red) attributed to the $^5D_0 \rightarrow {}^7F_2$ transition is used to reconstruct the 2D PL intensity maps in (b) for a 3 x 3 mm² area and, (c) for a 140 x 150 mm² area.

A further insight into the local environment of $Eu^{3+}$ ions in the crystallized spots (Fig. 5(a)) was obtained by collecting and analysing the PL intensity map (Fig. 5a) and the corresponding decay curve of the $^5D_0$ level under pulsed 532 nm laser excitation at room temperature (Fig. 5(b)). A single-exponential fit yields a lifetime of 1.8 ms for the $^5D_0$ level, comparable to lifetime values of a bulk reference sample reported in the literature [29] or measured under the same conditions in Fig. 5(b) (red curve). The slightly longer lifetime in the film is most likely due to a lower effective refractive index in the inhomogeneous crystallized spots (Fig. 5(a)) [31]. This result indicates that despite their non uniform localisation on the surface, $Eu^{3+}$ ions in the X2-YSO spots exhibit good optical properties, comparable to high-quality bulk crystals.

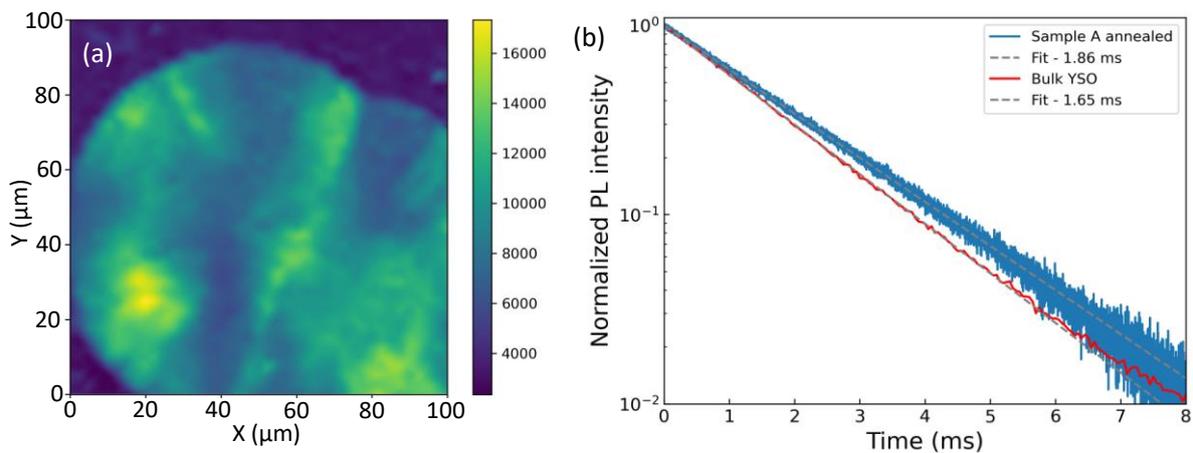

Fig. 5. (a) 100 x 100 mm² PL intensity map recorded at 612 nm using a custom microscopy system for sample A. The corresponding PL decay curve in (b) was measured at 612 nm ($^5D_0 \rightarrow {}^7F_2$) on the same spot. A single

exponential fit yields a population lifetime of 1.86 ms for the $^5D_0$ energy level in the film. The PL decay recorded from the bulk reference sample, yielding a lifetime of 1.65 ms, is also given for the sake of comparison.

4. **Discussion**

As previously mentioned, yttrium orthosilicate ($Y_2SiO_5$) exists in 2 different monoclinic polymorphs labelled X1 and X2 [32]. In our films, the high-temperature phase (X2), that has been reported to be stable at temperatures above 1200 °C is preferentially formed after annealing. On the other hand, while the disilicate ($Y_2Si_2O_7$) can exist in 5 different allotropic phases [33], after annealing our thin films, we mostly obtain the β phase reported to be stable above 1225 °C. In fact, our results indicate that, by tuning the initial Y/Si ratio, one can selectively obtain the desired crystalline phase after annealing. The spectral signature of $Eu^{3+}$ ions in this oxide matrix indeed acts as a powerful probe to identify phase transformation especially when the studied domains are limited laterally and in thickness. This would be particularly difficult to analyse using X-ray diffraction for example. It is also noteworthy that these YSO phases are formed at moderate annealing temperatures while crystallisation from the melt, the preferred fabrication technique to make bulk crystals of X2-YSO for quantum applications, requires temperatures as high 2000 °C.

The direct synthesis of crystalline YSO using DLI-CVD under our selected conditions was however unsuccessful (Fig. 1, left panel). Instead, $Eu^{3+}$ PL emission suggests the presence of small crystalline domains of $Y_2O_3$ that are likely to be embedded in a glassy $SiO_2$ matrix as schematized in Fig. 6. As the Y/Si ratio decreases, the proportion of these yttria clusters decreases with respect to the glass. Because $Eu^{3+}$ ions are known to be poor emitters in such an amorphous glass, the PL spectra show broad emissions with a lower intensity (as shown in Fig. 1(c) and Fig. 1(d)). Formation of the silicate phases after annealing at high temperature, could be explained by the diffusion of Si from the glassy matrix into the yttria clusters (Fig. 6, right panel). These domains that extend laterally are either of the ortho or disilicate phase depending on the starting film's composition (yttrium-rich or yttrium-poor). Ellipsometry measurements were performed on samples A and B before and after annealing and the results are shown in Fig. 7. Using a 2-layer model that includes an $SiO_2$ interface layer of varying thickness and a Cauchy equation for the top layer, we obtained reasonably good fits for this data. After annealing, oxygen diffusion from the surface and its reaction with the Si substrate led to a much thicker $SiO_2$ interfacial layer, from a few nm to about 150 nm, i.e. comparable to the thickness of the yttrium silicate film. This is consistent with recently reported results in which annealed films composition was measured by secondary ion mass spectroscopy (SIMS) [18]. In the opposite direction, Si diffusion from the substrate also likely occurred and contributed to enrich the silicate phase. This last phenomenon was particularly obvious when annealing sample A for longer times (10 h) at 1200°C or at higher temperatures, leading to the nucleation of a small proportion of parasitic phases of $Y_2Si_2O_7$.

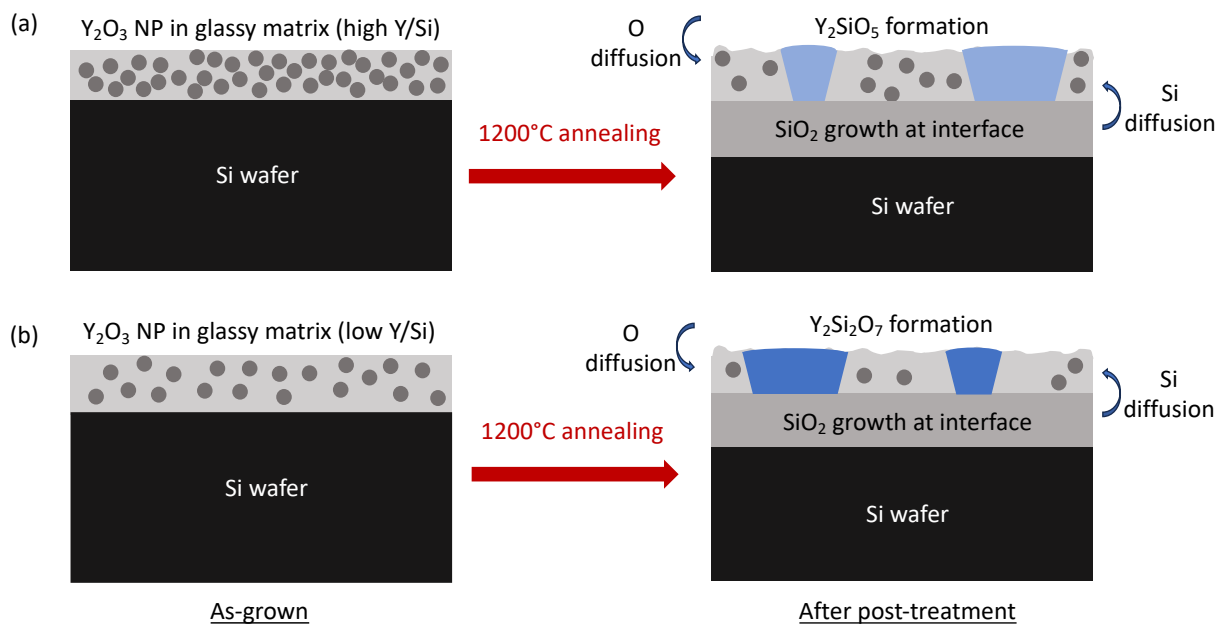

Fig. 6. Schematics illustrating the formation of both $Y_2SiO_5$ (a) and $Y_2Si_2O_7$ (b) after annealing. The initial thin film is composed of $Y_2O_3$ nanoclusters embedded in a glassy matrix (in grey) while diffusion leads to the appearance of the silicate phase (in blue).

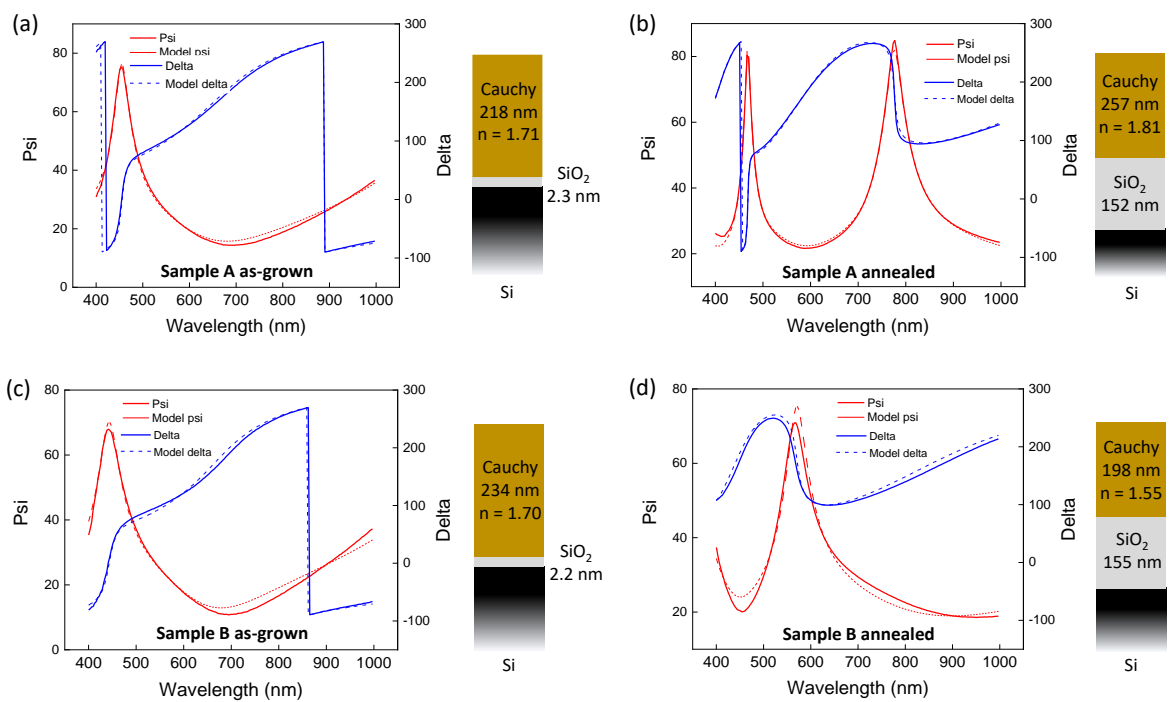

Fig. 7. Ellipsometric measurement of samples A and B as-grown (a) and (c) and after annealing (b) and (d). The changes in light polarization (psi and delta) are fitted with a model (dashed lines) that includes a $SiO_2$ interfacial

These experiments show that engineering a high-quality, single-crystalline and fully covering silicate film to host $Eu^{3+}$ ions requires appropriate deposition and annealing conditions. To do so, the use of other substrate materials or the deposition of much thicker films (several µm) may be a way to limit silicon diffusion into the film and oxygen to the interface. Fine tuning the annealing treatment, both in temperature, ramp and duration could also be explored to promote formation of the required phase over larger areas. Finally, direct formation of the silicate phase inside the growth chamber may be hampered by a low deposition temperature. It is currently limited to about 1000 °C in our set-up while temperatures above 1200 °C, better suited to favour the in-situ creation of X2-YSO, could be reached providing important technical modification of our heating system are made.

Despite the localized formation of the YSO phase in the CVD films, the promising optical properties exhibited by $Eu^{3+}$ ions in these areas, and in particular, the long population lifetime of the $^5D_0$ excited state (Fig. 4), indicate that even at the current status, these films could already be of interest for quantum technologies. Indeed, approaches involving addressing single ions through optical cavities as quantum computing schemes and single photon sources [7,34] do not necessarily require extended crystallized areas but ions embedded in a good local environment leading to good coherence properties i.e. long quantum-state lifetimes. Therefore, to fully confirm the potential of the YSO CVD thin films for quantum applications, a fundamental next step will be the assessment of the optical homogeneous and inhomogeneous linewidths of $Eu^{3+}$ ions in the crystallized spots using low temperature high-resolution spectroscopy techniques.

5. **Conclusion**

In this paper, we propose a new method to grow thin films of RE doped yttrium orthosilicate, a relevant material for the field of quantum technologies, on silicon wafers that offer prospects of scalability and integration. By varying the deposition conditions of our DLI-CVD system and the annealing post-treatment, we were able to engineer crystallized oxide films with the desired phase. In particular, we identified a set of parameters allowing for the obtention of the high-temperature phase (X2) of YSO. The mechanisms leading to the phase selection and that involves Si diffusion from a glassy matrix into surrounding yttria clusters is discussed. Finally, we evaluated the optical properties of $Eu^{3+}$ ions embedded in our films and found similar lifetimes for the $^5D_0$ level as compared to high-quality bulk

X2-YSO crystals obtained by conventional crystal growth methods. By refining this approach, it should be possible to obtain crystalline domains of YSO over larger areas and with a good phase purity. This thin-film fabrication approach opens interesting prospects into their use for realizing quantum devices that explore the coherent properties of RE ions.

**Ackowledgements:**

The CNRS Prime 80 grant (MATHYQ and Hyboxi) and the regional network on quantum technologies of Ile-de-France region (Quantip) are gratefully acknowledged for funding. This project has also received funding from the European Research Council (ERC) under the European Union's Horizon 2020 research and innovation programme (grant agreement n° 101019234, RareDiamond) and from the Quantum Flagship (grant agreement n° 820391, SQUARE). The authors would like to thank Hervé Guillon from Kemstream for help in selecting adapted precursors and deposition conditions for our DLI-CVD apparatus and would like to pay tribute to Dominique De Barros who significantly contributed to the development of the experiment.